\documentstyle[aas2pp4,psfig]{article}

\def\degree{{$^\circ$}}

\def\puncspace{\ifmmode\,\else{\ifcat.\C{\if.\C\else%
\if,\C\else\if?\C\else\if:\C\else\if;\C\else\if-\C\else%
\if)\C\else\if/\C\else\if]\C\else\if'\C%
\else\space\fi\fi\fi\fi\fi\fi\fi\fi\fi\fi}%
\else\if\empty\C\else\if\space\C\else\space\fi\fi\fi}\fi}%
\def\SP{\let\\=\empty\futurelet\C\puncspace}

\def\ee#1{\ifmmode {} \times 10^{#1} \else ${} \times 10^{#1}$\fi}
\def\sub#1{\ifmmode _{#1} \else $_{#1}$\fi}
\def\sup#1{\ifmmode ^{#1} \else $^{#1}$\fi}

\def\about{\ifmmode \sim \else {$\sim\,$}\fi}
\def\lta{\ifmmode {\,\mathbin{\lower 3pt\hbox   
    {$\,\rlap{\raise 5pt\hbox{$\char'074$}}\mathchar"7218\,$}}}
    \else {${\mathbin{\lower 3pt\hbox
    {$\rlap{\raise 5pt\hbox{$\char'074$}}\mathchar"7218\,$}}}
    $}\fi}
\def\gta{\ifmmode {\mathbin{\lower 3pt\hbox   
    {$\,\rlap{\raise 5pt\hbox{$\char'076$}}\mathchar"7218\,$}}}
    \else {${\mathbin{\lower 3pt\hbox
    {$\rlap{\raise 5pt\hbox{$\char'076$}}\mathchar"7218\,$}}}
    $}\fi}

\def\ast{\mathchar"2203} \mathcode`*="002A   

\def\degree{{\ifmmode ^\circ \else $^\circ$\fi}}

\def\Hz{{\hbox{Hz}}\SP}

\def\kHz{{\hbox{kHz}}\SP}

\def\aql#1{\leavevmode\hbox{Aql~X-#1}\SP}

\def\cyg#1{\leavevmode\hbox{Cyg~X-#1}\SP}
\def\fu#1{\leavevmode\hbox{4U~#1}\SP}
\def\gx#1{\leavevmode\hbox{GX~#1}\SP}

\def\ks#1{\leavevmode\hbox{KS~#1}\SP}

\def\sco#1{\leavevmode\hbox{Sco~X-#1}\SP}

\def\gro{\leavevmode{\it Compton Gamma-Ray Observatory\/}\SP}

\def\rxte{\leavevmode{\it RXTE\/}\SP}

\def\mdot{{\ifmmode \dot M \else {$\dot M$}\fi}}
\def\mdote{{\ifmmode \dot M_E \else {$\dot M_E$}\fi}}
\def\mdoti{{\ifmmode \dot M_i \else {$\dot M_i$}\fi}}
\def\msun{{\ifmmode M_\odot \else {$M_{\odot}$}\fi}}
\def\nonrot{{\rm 0}}

\begin{document}

\lefthead{Miller, Lamb, \& Cook}
\righthead{Neutron Star Equation of State Limits}

\title{Effects of Rapid Stellar Rotation on
Equation of State Constraints Derived from 
Quasi-Periodic Brightness Oscillations}

\author{M.\ Coleman Miller}
\affil{Department of Astronomy and Astrophysics, University of Chicago\\
       5640 South Ellis Avenue, Chicago, IL 60637, USA\\
       miller@bayes.uchicago.edu}
\authoremail{miller@bayes.uchicago.edu}

\author{Frederick K.\ Lamb}
\affil{Department of Physics and Department of Astronomy,
       University of Illinois at Urbana-Champaign\\
       1110 W. Green St., Urbana, IL 61801-3080, USA\\
       f-lamb@uiuc.edu}
\authoremail{f-lamb@uiuc.edu}

\author{Gregory B. Cook}
\affil{Center for Radiophysics and Space Research,
       Cornell University\\
       Ithaca, NY 14853\\
       cook@spacenet.tn.cornell.edu}
\authoremail{cook@spacenet.tn.cornell.edu}

\begin{abstract}

Quasi-periodic X-ray brightness
oscillations (QPOs) with frequencies
$\gta1$~kHz have now been discovered in
more than a dozen neutron stars in
low-mass X-ray binary systems using the
{\em Rossi X-ray Timing Explorer}. There
is strong evidence that the frequencies of
the kilohertz oscillations are the orbital
frequencies of accreting gas in nearly
circular orbits around these stars.
 Some stars that produce kilohertz QPOs
may have spin frequencies $\gta400$~Hz.
For spin rates this high, first-order
analytic treatments of the effects of the
star's rotation on its structure and the
spacetime are inaccurate. Here we use the
results of a large number of fully
relativistic, self-consistent numerical
calculations of the stellar structure of
rapidly rotating neutron stars and the
interior and exterior spacetime to
investigate the constraints on the
properties of such stars that can be
derived if stable circular orbits of
various frequencies are observed. We
have computed the equatorial radius of
the star, the radius of the innermost
stable circular orbit, and the frequency
of the highest-frequency stable circular
orbit as functions of the stellar spin
rate, for spin rates up to the maximum
possible and for several illustrative
equations of state.
 Our calculations show that the upper 
bounds on the stiffness of neutron star
matter implied by a given orbital frequency
are typically significantly stricter for
stars with spin frequencies $\gta 400$~Hz
than for slowly rotating stars.

\end{abstract}


\section{INTRODUCTION}

The successful launch of the {\em Rossi
X-ray Timing Explorer\/} (\rxte) has made
it possible to investigate, for the first
time, the X-ray variability of neutron
stars and black holes at frequencies
$\gta300$~Hz. One of the most important
discoveries made with \rxte is that many
neutron stars in low-mass X-ray binaries
produce high-frequency brightness
oscillations with frequencies in the range
$\sim$300--1200~Hz (see van der Klis
1998). High-frequency oscillations are
observed both during type~I
(thermonuclear) X-ray bursts and in the
persistent X-ray emission. The discovery
of these oscillations has made possible
derivation of interesting upper bounds on
the masses and radii of these neutron
stars and significant new constraints on
the equation of state of neutron star
matter (Miller, Lamb, \& Psaltis 1998a,
1998b; Lamb, Miller, \& Psaltis 1998;
Miller \& Lamb 1998; Strohmayer et al.\
1998).

Only a single oscillation has been
observed from each source during a type~I
X-ray burst, and the oscillations in the
tails of bursts appear to be highly
coherent (see, e.g., Smith, Morgan, \&
Bradt 1997), with frequencies that are
always the same in a given source (see,
e.g., Strohmayer 1997). The
burst oscillations are thought to be
caused by the existence of one or two
brighter regions on the stellar surface
that produce oscillations at the stellar
spin frequency or its first overtone as
the star spins (see Strohmayer et al.\
1997b for compelling evidence in favor of
this interpretation). The frequency of
the burst oscillations ranges from
$\sim$330~Hz in \fu{1702$-$42} (Swank
1997) to 589~Hz in an unidentified source
in the direction of the Galactic center
(Strohmayer et al.\ 1997a).

The kilohertz quasi-periodic oscillations
(QPOs) observed in the persistent emission
have high amplitudes and relatively high
coherences. The frequencies of the two
QPOs often observed simultaneously in a
given source have a frequency separation 
that is almost constant in
all sources (see Wijnands \& van der Klis
1997; Psaltis et al.\ 1998a, 1998b), although
the frequencies of the QPOs themselves
vary by hundreds of Hertz (see Wijnands et
al.\ 1998 and van der Klis 1995). The separation
frequencies of the two kilohertz QPOs seen
in \fu{1728$-$34} (Strohmayer et al.\ 1996,
1997b) and \fu{1702$-$42}
(Swank 1997) are consistent with the
frequencies of their burst oscillations.
The separation frequencies in
\fu{1636$-$536} (Wijnands et al.\ 1997;
Zhang et al.\ 1997) and \ks{1731$-$260}
(Smith et al.\ 1997; Wijnands \& van der
Klis 1997) are consistent with half the
frequencies of their burst oscillations.

The presence of only two simultaneous
kilohertz QPOs in a given source, the
approximately constant frequency
separation $\Delta\nu$ between them, and
the consistency of $\Delta\nu$ with the
stellar spin frequency inferred from
burst oscillations is strong evidence
that the stellar spin is generating the
frequency difference, that only one
sideband of the primary QPO frequency is
being generated, and that one of the two
QPOs is therefore caused by the beat of
the spin frequency with the other
frequency (see Lamb et al.\ 1998; Miller
et al.\ 1998a). This implies that in
addition to the spin frequency there is
only one other primary frequency and
that this frequency is a rotational
frequency such as an orbital frequency. 
This excludes neutron star
surface and photon bubble oscillations as
explanations for the primary kilohertz
QPO frequency and makes disk
oscillations an improbable explanation
(see van der Klis 1998; Lamb et al.\
1998).

In all sources the frequencies of the
kilohertz QPOs fall within the expected
range of orbital frequencies near a
neutron star and can vary by several
hundred Hertz in less than a few hundred
seconds while remaining highly coherent
($\nu/\Delta\nu \sim 100$). This is
strong further evidence against disk
oscillations and in favor of orbital
motion of inhomogeneities in the
accretion disk as the cause of the
primary kilohertz QPO (see Lamb et al.\
1998). The accreting gas exerts a strong
torque on the neutron star and hence the
star is expected to be spinning in the
same sense as the orbital motion of the
accreting gas. The beat frequency must
therefore be the lower of the two
kilohertz QPO frequencies whereas the
orbital frequency is the higher.

There are two candidates for the
orbital frequency: the frequency at the
radius where the accreting gas first
couples strongly to magnetic field of
the neutron star and the frequency at
the sonic radius where radiation forces
or general relativistic effects cause the
radial motion of the gas to increase
sharply and become supersonic. The
general properties of the kilohertz QPO
sources and the specific properties of
the kilohertz QPOs themselves strongly
indicate that the relevant frequency is
the orbital frequency at the sonic point
(Miller et al.\ 1998a; see also Lamb et
al.\ 1998). In either case, the
frequency $\nu_{\rm QPO2}$ of the
higher-frequency kilohertz QPO is the
orbital frequency of gas in a nearly
circular orbit around the neutron star,
whereas the frequency $\nu_{\rm QPO1}$
of the lower-frequency QPO is the beat
of the neutron star spin frequency with
an orbital frequency near $\nu_{\rm
QPO2}$.

In the following discussion we shall
describe a circular orbit as stable or
unstable according to its properties as
determined by solving the geodesic
equation for a test particle moving in
that orbit in the spacetime of interest. However, 
it is important to bear in mind that {\em there
are no closed, circular orbits in the
vicinity of an accreting neutron star},
because the motion of gas near such a
star is affected not only by the
curvature of spacetime but also by
radiation, magnetic, and viscous forces,
which cause the gas to spiral inward
even at radii where, in their absence,
closed, stable orbits would be possible
(Miller \& Lamb 1993, 1996). However,
the distinction between stable circular
orbits (SCOs) and unstable circular
orbits is still relevant. In particular,
the innermost stable circular orbit
(ISCO) is still physically significant
when the effects of radial gas pressure
forces on the ISCO can be neglected
(which should be valid when the
luminosity of the source is much less
than the Eddington critical luminosity),
because under these conditions gas {\em
inside\/} the ISCO spirals inward so
quickly that it cannot produce a
wavetrain with the coherence observed
for the kilohertz QPOs, regardless of
whether it is acted on by radiation,
magnetic, and viscous forces (Miller et
al.\ 1998a).

Identification of the
higher-frequency kilohertz QPO with the
frequency of an SCO has made it possible to
derive upper bounds on the masses and
radii of the neutron stars in the
kilohertz QPO systems (Miller et al.\
1998a, 1998b). These bounds follow from the
requirement that the radius $R_{\rm orb}$ of 
the orbit be larger than the radius $R_{\rm ms}$ 
of the ISCO as well as larger than the
equatorial radius $R_{\rm eq}$ of the star
(if $R_{\rm eq} < R_{\rm orb} < R_{\rm
ms}$, orbits with the required frequency
exist but are unstable).

For nonrotating stars, observation of a
given orbital frequency can be used to
derive upper
bounds on the mass and radius that are
independent of the equation of state
assumed (Miller et al.\ 1998a). For
rotating stars, the situation is more
complicated. In general, both the
structure of the star and the spacetime
are affected by the star's rotation, and
there are no general analytical
expressions for the relevant quantities.
However, the exterior spacetime of a
slowly and uniformly rotating fluid star is unique to
first order in the dimensionless angular
momentum $j\equiv cJ/GM^2$, where $J$ and
$M$ are the star's angular momentum and
gravitational mass, and can be expressed
analytically to this order (Hartle \&
Thorne 1968). The leading corrections to
the expressions for the orbital frequency
and the radius of the ISCO are
first-order in $j$. Using these
expressions, one can demonstrate that
observation of a given orbital frequency
also implies upper bounds on the mass and
radius of a slowly rotating star (Miller
et al.\ 1998a). For a given stellar spin
frequency, these upper bounds depend on
the moment of inertia and hence on the
equation of state assumed.

Many of the kilohertz QPO sources appear
to have spin frequencies $\sim$250--350~Hz
(see Miller et al.\ 1998a). Examples of
such sources include \fu{0614$+$091},
\fu{1608$-$52}, \fu{1820$-$30}, \cyg2,
\sco1, \gx{5$-$1}, and \gx{17$+$2}, all of
which have kilohertz QPO separation
frequencies in this range, as well as
\fu{1728$-$34} and \fu{1702$-$42}, which
not only have kilohertz QPO separation
frequencies in this range but also have
burst oscillation frequencies that are
consistent with these separation
frequencies. The $\sim$520 and
$\sim$580~Hz frequencies of the burst
oscillations seen in \ks{1731$-$260} and
\fu{1636$-$536} are thought to be twice
their spin frequencies, although this is
not certain. For spin frequencies in
this range, $j$ is $\sim$0.1--0.3,
depending on the assumed equation of
state and the mass of the star, and hence
an analysis that is first-order in $j$ is
quite accurate for such stars. Such an
analysis shows that spin rates
$\sim$300~Hz can increase the upper bound
on the stellar mass by as much as $\sim$10--20\%
but typically increase the upper bound
on the radius by only $\sim$2--5\%
(Miller et al.\ 1998a, 1998b).

On the other hand, some neutron stars
that show kilohertz QPOs may turn out to
have spin frequencies $\gta400$~Hz. For
example, oscillations with frequencies
$\sim$550 and $\sim$590~Hz have been seen
during X-ray bursts from, respectively,
\aql1 and the unknown source in the
direction of the galactic center,
possibly indicating that these neutron
stars have spin frequencies this high
(Miller et al.\ 1998a). The recent 
discovery using \rxte that the source
SAX J1808.4--3658
has a coherent 401~Hz oscillation indicates
that this accreting neutron star is spinning
rapidly
(Wijnands \& van der Klis 1998a, 1998b). For stars
spinning this fast, the effect of the
star's spin on its equilibrium structure
(which is second-order) and on the
spacetime can be substantial. In order
to obtain accurate results for such high
spin rates, the equilibrium stellar
structure and the interior and exterior
spacetime must be computed
self-consistently, which can only be
done numerically.

Here we use the results of a large number
of fully relativistic, self-consistent
numerical calculations of the structure
of rapidly rotating neutron stars and
the interior and exterior spacetime to
investigate the constraints on the
properties of such stars that can be
derived if SCOs of various frequencies
are observed. We have computed the
equatorial radius of the star, the
radius of the ISCO, and the frequency of
the highest-frequency SCO as functions
of the stellar spin rate and gravitational
mass, for spin rates
up to the maximum possible and for
several illustrative equations of state.
Comparison of these results with the
highest observed kilohertz QPO frequency
in a given source can be used to derive
bounds on the mass and radius of the
neutron star in that source, for a given
equation of state. We also report the
frequency of the highest-frequency SCO
as a function of the stellar spin rate,
for stars of any mass constructed using
a given equation of state. These curves
can be used to check whether a particular
equation of state is consistent with the
frequency of a given kilohertz QPO.

Our calculations show that the
upper bounds on the stiffness of
neutron star matter implied by a given
orbital frequency are typically significantly
stricter for stars with spin frequencies
$\gta400$~Hz than for slowly rotating
stars. 

In \S~2 we describe our assumptions and
calculational methods. In \S~3 we
present our results and discuss the
implications for constraining the
properties of neutron star matter. Our
conclusions are summarized in \S~4.

\section{ASSUMPTIONS AND METHOD}

In deriving bounds on the masses and
radii of the neutron stars with
kilohertz QPOs, we assume that the
higher-frequency of the two simultaneous
kilohertz QPOs is the frequency of a
stable circular orbit around the neutron
star, for the reasons discussed in \S~1. 
Hence the orbital radius $R_{\rm
orb}$ that corresponds to the QPO
frequency must be larger than both the
equatorial radius $R_{\rm eq}$ of the
neutron star and the radius $R_{\rm ms}$
of the ISCO (Miller et al.\ 1998a).

We compute the equilibrium stellar
structure and the interior and exterior
spacetime using the numerical code
described in Cook, Shapiro, \& Teukolsky
(1992, 1994a, 1994b). This code solves
the full general relativistic equation of
hydrostatic equilibrium for a star with a
given spin rate using a variation of the
metric potential method of Komatsu,
Eriguchi, \& Hachisu (1989a, 1989b). It
gives accurate solutions even for stars
that are spinning very rapidly.

In any stationary, axisymmetric
spacetime, the orbital frequency at a
given coordinate radius, as measured at infinity, is
 $\Omega={d\phi}/{dt}$,
 where $\phi$ and $t$ are, respectively,
the global azimuthal and time coordinates
based on the spacelike and timelike
Killing vector fields of the spacetime.
The time interval required for one orbit
of an element of gas is the same
everywhere, as measured in the global
time coordinate. Given the metric of the
exterior spacetime, the orbital frequency
at a given radius is the solution of the
equation (see Lightman et al.\ 1973, p.\
469)
 \begin{equation}
 g_{\phi\phi,r}\Omega^2+2g_{t\phi,r}\Omega+g_{tt,r}=0\; ,
 \end{equation}
 where $g_{\phi\phi}$, $g_{t\phi}$, and
$g_{tt}$ are the metric components
indicated and commas denote partial
derivatives.

\subsection{Masses and Equations of State}

We have explored the constraints implied
by observation of an SCO of given frequency
for a variety of neutron star equations of
state. These restrictions are most
significant if the equation of state is
hard rather than soft. Hence, in this
report we present results for four
relatively hard equations of state.
For completeness, we consider both
baryonic masses that are stable for
nonrotating stars (the so-called ``normal"
sequences of Cook et al.\ [1994b])
and the higher baryonic masses that 
are stable only for rotating stars (the
``supramassive" sequences of Cook et 
al.\ [1994b]). Whether the supramassive sequences
are accessible depends on how the
specific angular momentum of
the accreting gas evolves with time.

In order to facilitate comparisons with
previous studies of neutron star
properties (see, e.g., Pethick \&
Ravenhall 1995), we consider the
Friedman-Pandhari\-pande-Skyrme (FPS)
equation of state (Friedman \&
Pandharipande 1981; Lorenz, Ravenhall, \&
Pethick 1993). The FPS equation of state
is based on the Urbana $v_{14}$
two-nucleon potential plus the
density-dependent three-nucleon
interaction model of Lagaris \&
Pandharipande (1981) and gives a maximum
gravitational mass for a nonrotating star of about
$1.8\,\msun$, compared with a maximum mass of
$2.12\,\msun$ for a rotating star. The maximum spin 
frequency for stars in the normal sequence is 1411~Hz, 
and the maximum spin frequency for stars
in the supramassive sequence is 1878~Hz.

As an example of later realistic equations
of state, we consider the UU equation of
state (Wiringa, Fiks, \& Fabrocini 1988),
which is based on the Urbana $v_{14}$
two-nucleon potential plus the Urbana VII
three-nucleon potential (Schiavilla,
Pandharipande, \& Wiringa 1986) and gives
a maximum mass for a nonrotating star of
about $2.2\msun$. Although it is based on
older scattering data, the UU equation of
state is similar to the recent ${\rm A18}
+ {\rm UIX'} + \delta v_{\rm b}$ equation
of state (Akmal, Pandharipande, \& Ravenhall
1998), which is based on the modern
Argonne $v_{18}$ two-nucleon potential
and the Urbana IX three-nucleon
potential and takes into account the
nonzero momentum of the interacting
nucleons (see Pandharipande, Akmal, \&
Ravenhall 1998). Like
the ${\rm A18} + {\rm UIX'} +
\delta v_{\rm b}$ equation of state, the
UU equation of state gives
a maximum mass of about $2.2\,\msun$ for
a nonrotating neutron star. The maximum
mass for a rotating neutron star is
$2.61\,\msun$, and the maximum rotation
frequencies for the normal and supramassive
sequences are, respectively, 1561~Hz and
1989~Hz.

In order to illustrate the generic
effects of significant softening of a
stiff equation of state at a critical
density, we consider
the tensor interaction (TI) equation of
state of Pandharipande \& Smith (1975a; M
in the Arnett \& Bowers [1977] survey).
Although the TI equation of state is
itself no longer of interest to nuclear
physicists, this equation of state
demonstrates the effects of a very
strong, first-order phase transition,
such as may occur at the transition from
nucleon matter to quark matter (see
Glendenning 1992; Heiselberg, Pethick, \&
Staubo 1993; Pandharipande et al.\ 1998).
The maximum mass of a nonrotating star
constructed using the TI equation of
state is $1.8\,\msun$, and the maximum
mass of a rotating star is $2.1\,\msun$.
The maximum rotation frequency for the
normal sequence is 707~Hz, and for the
supramassive sequence is 1229~Hz.

Finally, as an example of the relatively
stiff equations of state often given by
mean field theories, we consider the
mean-field equation of state of
Pandharipande \& Smith (1975b; L in the
Arnett \& Bowers [1977] survey). The
maximum mass of a nonrotating star
constructed using this equation of state
is $2.7\msun$, compared with 
$3.27\msun$ for a rotating star. The maximum rotation 
frequency is 1031~Hz for the normal
sequence and 1321~Hz for the supramassive
sequence.

\subsection{First-Order Expressions}

In \S~3 it will be instructive to compare
the behavior of the orbital frequencies
and radii computed using our numerical
models of rapidly rotating stars with the
behavior given
by the analytical expressions valid for
slowly rotating stars. As noted in \S~1,
the spacetime around a rotating fluid
star is unique to first order in the
dimensionless angular momentum $j$. To
this order in $j$, the frequency of a
prograde orbit at circumferential radius
$r$ around a star with gravitational mass
$M$ is (see Lightman et al.\ 1973, p. 469; 
Miller et al.\ 1998a)
 \begin{equation}
 \label{Omega}
 \Omega=\left[1-j(M/r)^{3/2}\right]
 \left(M/{r^3}\right)^{1/2}
 \end{equation}
 and the circumferential radius of the
ISCO is
 \begin{equation}
 \label{Rms}
  R_{\rm ms}(M,j) \approx
   6M[1-j(2/3)^{3/2}]\; ,
 \end{equation}
 in units in which \hbox{$G \equiv c \equiv 1$}.
In the present work we always quote
circumferential radii, in contrast to
Miller et al.\ (1998a), where we quoted
Boyer-Lindquist radii. The two radii are identical
to first order in $j$, but to
higher orders in $j$ the circumferential
radius is the physically meaningful
radius,  which is the reason we use it
here.
 Combining equations~(\ref{Omega})
and~(\ref{Rms}), one can show that to
first order in $j$, the frequency
$\nu_{\rm K, ms}$ of the innermost stable
prograde orbit is (see Klu\'zniak,
Michelson, \& Wagoner 1990; Miller et al.\
1998a)
 \begin{equation}
 \label{nuKms}
 \nu_{\rm K, ms} \approx
   2210\,(1+0.75j)(M_\odot/M)\,\Hz\;.
 \end{equation}
 Thus, for slowly rotating stars the
frequency of the ISCO {\em increases
linearly\/} with the star's spin rate. 

Using
expressions~(\ref{Omega})--(\ref{nuKms}),
one can show (Miller et al.\ 1998a) that
the mass and radius of a slowly rotating
star are bounded above by
 \begin{equation}
 \label{Mmax}
   M_{\rm max} \approx
   [1+0.75j(\nu_{\rm spin})]M^\nonrot_{\rm max}
 \end{equation}
 and
 \begin{equation}
 \label{Rmax}
   R_{\rm max} \approx
   [1+0.20j(\nu_{\rm spin})]
   R^\nonrot_{\rm max}\;.
 \end{equation}
 Here $j(\nu_{\rm spin})$ is the value
of $j$ for the observed stellar spin rate
at the maximum allowed mass for the
equation of state being considered and
 \begin{equation}
 \label{MmaxO}
 M^0_{\rm max} = 2.2\,
 (1.0~\kHz/\nu^\ast_{\rm QPO2})~M_\odot
 \end{equation}
 and
 \begin{equation}
 \label{RmaxO}
 R^0_{\rm max} = 19.5\,
 (1.0~\kHz/\nu^\ast_{\rm QPO2})~{\rm km}
 \end{equation}
 are the upper bounds on the mass and
radius of a nonrotating star in terms of
$\nu^\ast_{\rm QPO2}$, the highest
observed frequency of the
higher-frequency kilohertz QPO.
 The precise upper bounds on the mass and
radius depend on the equation of state
and can be determined by searching a grid
of neutron star models for the one that
gives the maximum allowed mass.
Equations~(\ref{Mmax}) and~(\ref{Rmax})
show that the bounds are always greater
for a slowly rotating star than for a
nonrotating star, regardless of the
equation of state assumed. 

No expressions similar to
equations~(\ref{Omega})--(\ref{Rmax}) are
available for rapidly rotating stars. 

\section{RESULTS AND DISCUSSION}

We first show how the radius of the ISCO
and the equatorial radius
vary with stellar spin rate for stars
constructed using the FPS equation of
state. The
behavior of these radii makes clear why
the frequency of the highest-frequency
SCO around a star of given mass generally
first increases as the star is spun up
and then decreases. Considering this
behavior for stable stars with different
masses makes the behavior of the maximum
frequency of an SCO for stars of any mass
understandable.

Next, we present mass-radius relations for
stars with a wide range of spin rates,
constructed using the FPS and UU
equations of state. We then show how to
derive limits on the mass and radius of a
rapidly rotating star from the frequency
of an SCO around it and discuss the
constraints on the equation of state of
neutron star matter implied by such
constraints.

\subsection{Radii and Orbital
Frequencies}

Figures 1a and 1b show how the circumferential radius of the ISCO
and the circumferential radius of the stellar equator vary with 
$\nu_{\rm spin}$, the
stellar spin frequency measured at infinity, for
stars constructed using the FPS equation of state. These stars have
constant baryonic masses equal to
those of non-rotating stars with 
gravitational masses of $1.4M_\odot$ and $1.6M_\odot$.
Because the gravitational mass increases only slightly with
increasing spin frequency, the curves for stars of constant
gravitational mass are almost identical (the largest stable
equatorial radii are very slightly smaller).
As expected, the
dimensionless angular momentum $j$
increases linearly with spin rate for
slowly rotating stars but more steeply
for rapidly rotating stars: for the
$1.4\,M_\odot$ model, \hbox{$j=0.23$} at
\hbox{$\nu_{\rm spin}=0.5$~kHz} and 0.52
at 1.0~kHz; for the $1.6\,M_\odot$
models \hbox{$j=0.20$} at 0.5~kHz and
0.43 at 1.0~kHz; for the $1.8\,M_\odot$
models \hbox{$j=0.16$} at 0.5~kHz and
0.36 at 1.0~kHz. Figure~1c shows how the
frequency of the highest-frequency SCO
varies with stellar spin rate of
$1.4\,M_\odot$, $1.6\,M_\odot$, and
$1.8\,M_\odot$ stars.

 \begin{figure}
 \centering
 \vglue-0.35truein
 \hbox{\hskip 0.0truein
\psfig{file=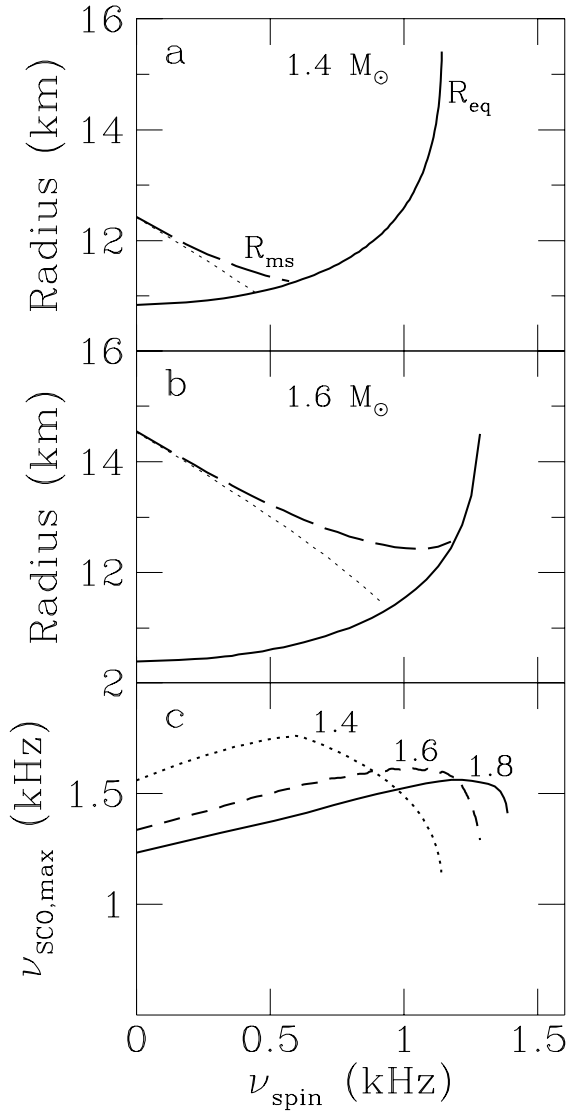,width=6.4truein}}
 \vglue-0.1truein
 \caption[fig1]{
 Typical variations of important radii
and frequencies with stellar spin rate.
 (a)~Circumferential radii of the
innermost stable circular orbit (dashed
line) and the stellar equator (solid
line) as a function of spin rate, for a 
1.4$M_\odot$ star. Also shown for
comparison is the circumferential radius
of the innermost stable circular orbit in
a Kerr spacetime with the same
gravitational mass and angular momentum
(dotted line).
 (b)~Same radii as in (a), but for a
1.6$M_\odot$ star.
 (c)~Frequency of the highest-frequency
stable circular orbit as a function of
stellar spin rate, for $1.4\,M_\odot$
(dotted line), $1.6\,M_\odot$ (dashed
line), and $1.8\,M_\odot$ (solid dashed
line) stars.
 All stellar models were constructed
using the FPS equation of state.
 }
 \end{figure}

The circumferential radius $R_{\rm ms}$
of the ISCO decreases linearly with spin
rate for slowly rotating stars, in
agreement with the first-order expression
(\ref{Rms}), but decreases more slowly as
the spin rate increases. For the
$1.4\,M_\odot$ star, the deviation from
equation~(\ref{Rms}) is significant at
$\nu_{\rm spin} \approx 300$~Hz (see
Fig.~1a). For the $1.6\,M_\odot$ star, the
deviation is significant at 500~Hz and at
about 1065~Hz $R_{\rm ms}$ reaches a
minimum and then {\em increases\/} with
increasing spin rate, until an ISCO no
longer exists (see Fig.~1b). In contrast,
the circumferential equatorial radius of
the stellar models increases quadratically
with the spin rate from \hbox{$\nu_{\rm
spin} = 0$}, exceeding $R_{\rm ms}$ at
about 580~Hz for the $1.4\,M_\odot$ star
and at about 1220~Hz for the
$1.6\,M_\odot$ star. For spin rates
above these critical rates, all circular
orbits with radii larger than the star's
equatorial radius are stable: there is no
ISCO.

The Kerr spacetime can be expressed
analytically and is therefore sometimes
used as a convenient approximation to the
exterior spacetime of a spinning neutron
star. For this reason, in Figures~1a
and~1b we compare $R_{\rm ms}({\rm
Kerr})$, the circumferential radius of
the ISCO in a Kerr spacetime with the
same gravitational mass and angular
momentum as the stellar models, with the
actual radius $R_{\rm ms}$ of the ISCO.
Unlike the actual radius, $R_{\rm
ms}({\rm Kerr})$ decreases monotonically
and nearly linearly with spin rate even
at high spin frequencies. Indeed, at
high spin rates $R_{\rm ms}({\rm Kerr})$
decreases {\em faster\/} than linearly
with increasing spin rate, and hence the
exact Kerr expression for $R_{\rm ms}$ is
a worse approximation than the
first-order approximation~(3). 

$R_{\rm ms}({\rm Kerr})$ is
significantly smaller than $R_{\rm ms}$
at high spin rates. As a result, when
$R_{\rm ms}$ is larger than the
equatorial radius of the star, the
frequency of the highest-frequency SCO
is significantly {\em lower\/} than one
would estimate using the Kerr spacetime,
and the constraints on the mass and
radius of the star are correspondingly
tighter. For both the 1.4$M_\odot$ and
1.6$M_\odot$ stars, the critical spin
rate at which the ISCO disappears in
the Kerr approximation is about 23\%
smaller than in the actual spacetime.
Thus, for stellar spin rates
$\gta400$~Hz, the exterior spacetime of
a spinning black hole is generally an
inaccurate approximation to the exterior
spacetime of a neutron star.

Figure~1c shows why the constraints on
the mass and radius of a slowly rotating
star implied by a given SCO frequency
are generally looser for a slowly
rotating star than for a nonrotating
star of the same mass, whereas the
constraints on a rapidly rotating star
are usually much tighter. For slowly
rotating stars with gravitational masses
of 1.4, 1.6, and $1.8\,M_\odot$
constructed using the FPS equation of
state, the equatorial radius of the star
is smaller than the radius of the ISCO.
Hence, at low spin rates the
highest-frequency SCO is the ISCO,
which at these spin rates shrinks
linearly as the spin rate increases (see
eq.~[\ref{Rms}]), causing the frequency
of the highest-frequency SCO to {\em
increase\/} linearly with the star's spin
rate. However, at a certain critical spin
rate the equatorial radius of the star
becomes larger than the radius of the
ISCO for a star with the given
gravitational mass and angular momentum.
For spin rates above this critical spin
rate, the highest-frequency SCO is the
orbit that just skims the stellar
surface. At high spin rates, the
equatorial radius of the star increases
rapidly with increasing spin rate,
causing the frequency of the
highest-frequency SCO to {\em
decrease\/} rapidly.

For the \hbox{$M=1.4\,M_\odot$} star, the
frequency of the highest-frequency SCO is
maximized at the spin frequency for which
the radius of the ISCO is equal to the
radius of the stellar equator (see
Fig.~1a). However, the highest-frequency
SCO for a star of given mass but any
possible spin rate is not necessarily the
ISCO with radius equal to the equatorial
radius of the star. This is illustrated
by the \hbox{$M=1.6\,M_\odot$} star. For 
this star, the frequency of the
highest-frequency SCO has its maximum at the
spin frequency at which the radius of the
ISCO is a minimum (see Fig.~1b). At this
frequency the radius of the ISCO is larger 
than the equatorial radius of the star. 

\subsection{Maximum SCO Frequency}

The maximum frequency of the
highest-frequency SCO for stable stars of
{\em any\/} mass can be determined by
constructing curves like those shown in
Figure~1c, for a dense sequence of
stellar masses. The curve of maximum
frequency as a function of spin rate is
then the upper envelope of these curves.
Figure~2 shows curves of maximum SCO
frequency as a function of stellar spin
rate, for stable stars of {\em any\/}
mass constructed with the four
indicated equations of state. If the
measured spin frequency of the star and
the frequency of a nearly circular orbit
correspond to a point that lies above
the curve for a given equation of state,
that equation of state is excluded
for all neutron stars.

 \begin{figure}[t!]   
 \centering
 \vglue-0.0truein
 \hbox{\hskip 0.0truein
\psfig{file=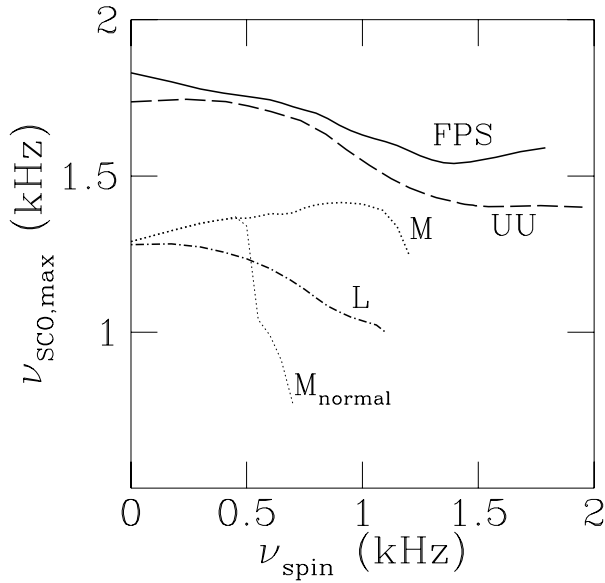,width=3.25truein}}
 \caption[fig2]{
 Maximum frequency of a stable circular
orbit as a function of spin rate for
stable stars of {\em any\/} mass, for the
four indicated equations of state, which
are discussed in the text. The curve labeled
``M$_{\rm normal}$" includes only normal sequences,
whereas the other curves include both normal
and supramassive sequences (see \S~3.1 for a 
discussion). 
 }
 \end{figure}

As Figure~2 shows, the maximum frequency
of the highest-frequency SCO {\em for
stars of any mass\/} typically decreases
with increasing spin rate, even though
the frequency of the highest-frequency
SCO {\em for a fixed gravitational
mass\/} increases with increasing spin
rate over a wide range of spin rates.
The reason is that at low spin rates the
mass that gives the highest-frequency
SCO is the mass at which the ISCO
coincides with the stellar equator. This
mass increases with increasing spin
rate, causing the maximum frequency to
decrease.

Stars constructed with the TI equation of
state M are an exception to this rule.
This equation of state is extremely stiff
at the low densities characteristic of
the centers of lower-mass stars, but
becomes very soft at the density reached
at the center of a nonrotating $M \approx
1.75\,M_\odot$ star, because a pion
condensate forms at this density. For
this equation of state, the
maximum orbital frequency occurs for a
stellar mass near the maximum mass
allowed by the star's spin frequency
(e.g., $1.8\msun$ for a slowly rotating
star).

Consider first the normal sequences.
At low spin frequencies, the surface of
the $1.8\,M_\odot$ star is well inside
the ISCO, the highest-frequency SCO is
therefore the ISCO, and the frequency of
the maximum-frequency SCO therefore
{\em increases\/} with increasing spin
rate. However, at about 500~Hz the
maximum frequency stops increasing and
then plummets, as shown by the curve
labeled ``M$_{\rm normal}$" in Figure~2. 
The reason is a general
relativistic effect first pointed out by
Cook et  al.\ (1994b): as the angular
momentum of a star increases, the spin
frequency first
increases, then decreases, and finally
increases again, producing a local
maximum in the spin frequency vs.\
angular momentum relation. For normal
sequences using equation
of state M, this local maximum occurs at
a spin frequency slightly greater than
500~Hz. Hence, in order to have an
observed spin frequency higher than
this, the star must have a much higher
angular momentum, but a star with this
much angular momentum has a much larger
equatorial radius, larger than the ISCO
for a star of its mass and angular
momentum, so the highest-frequency SCO
is at the stellar surface and has a
smaller frequency. 

Consider now the supramassive sequences.
For a fixed baryonic mass, there is a local
maximum in the spin frequency vs.\ angular
momentum relation, just as for the normal
sequences. A curve of $\nu_{\rm SCO,max}$ vs.\ 
$\nu_{\rm spin}$ constructed for a supramassive
sequence with a fixed baryonic mass would
therefore look similar to the ``M$_{\rm normal}$" curve in
Figure~2, except that it would start at a
positive spin frequency and would plummet at
a higher spin frequency. The local maximum of
the spin frequency increases with 
increasing baryonic mass. The envelope of
these curves produces the curve labeled 
``M". The rapid downturn of this curve at
a spin frequency of approximately 1100~Hz
occurs because above this frequency there is
no ISCO, and consequently the highest-frequency orbit
is the one that skims the surface. 
For $\nu_{\rm spin}>1100$~Hz, the equatorial
radius increases rapidly with increasing spin
frequency, and therefore the maximum
frequency of a circular orbit decreases rapidly.

Comparison of the two curves in 
Figure~2 for equation of state M demonstrates
the potential importance of the supramassive
sequences. For example, if only normal sequences 
are physically accessible (e.g., if the gas
accretes with low specific angular momentum),
then observation of an 1100~Hz SCO from a
star with a spin frequency in excess of
550~Hz would rule out equation of state M,
whereas equation of state M would still be
viable if supramassive sequences are
accessible.

 \begin{figure*}[t!]   
 \centering
 \vglue-0.1truein
 \hbox{
 \hskip -0.1truein
\psfig{file=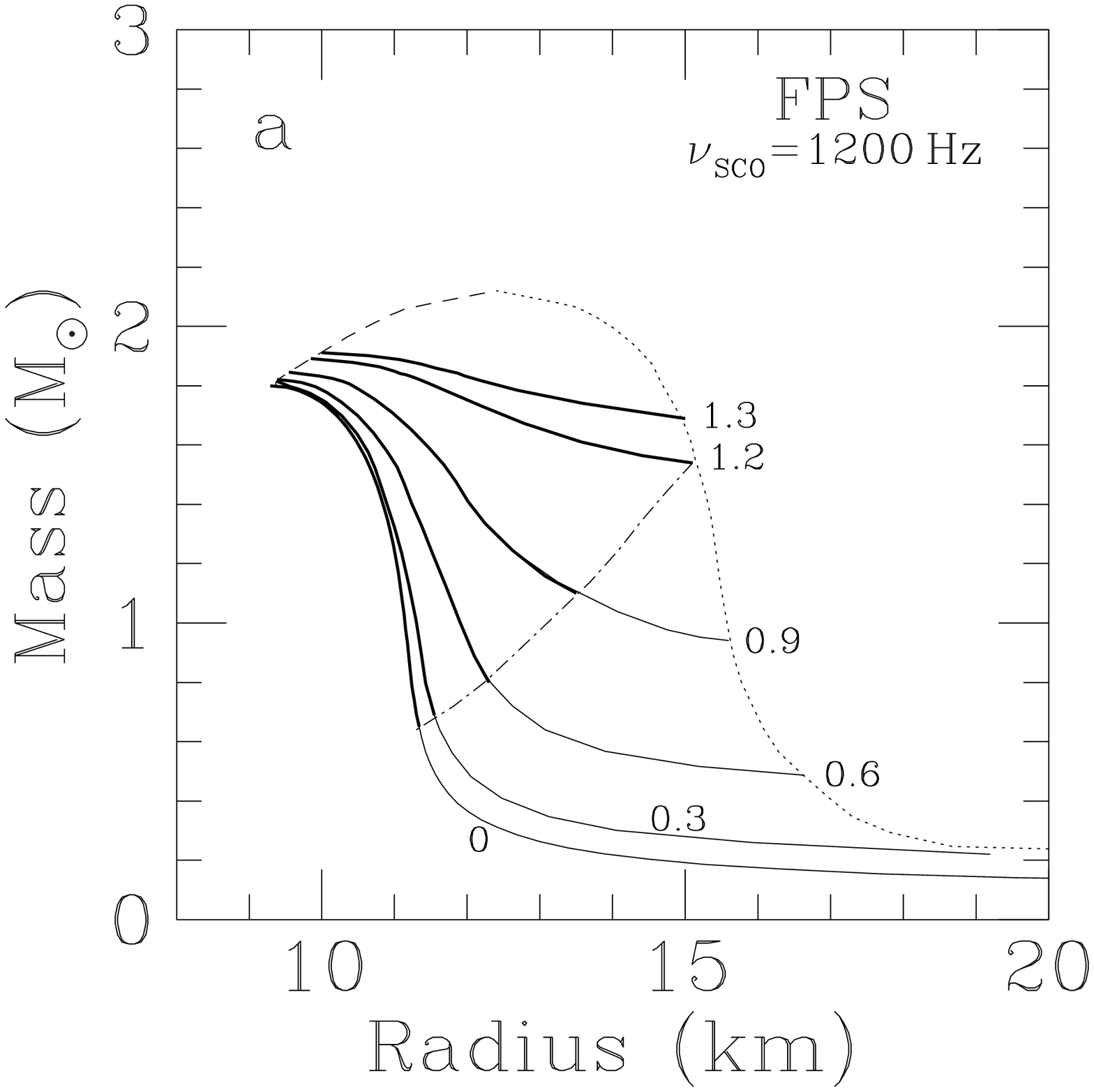,width=3.55truein}
\hskip -0.2truein
\psfig{file=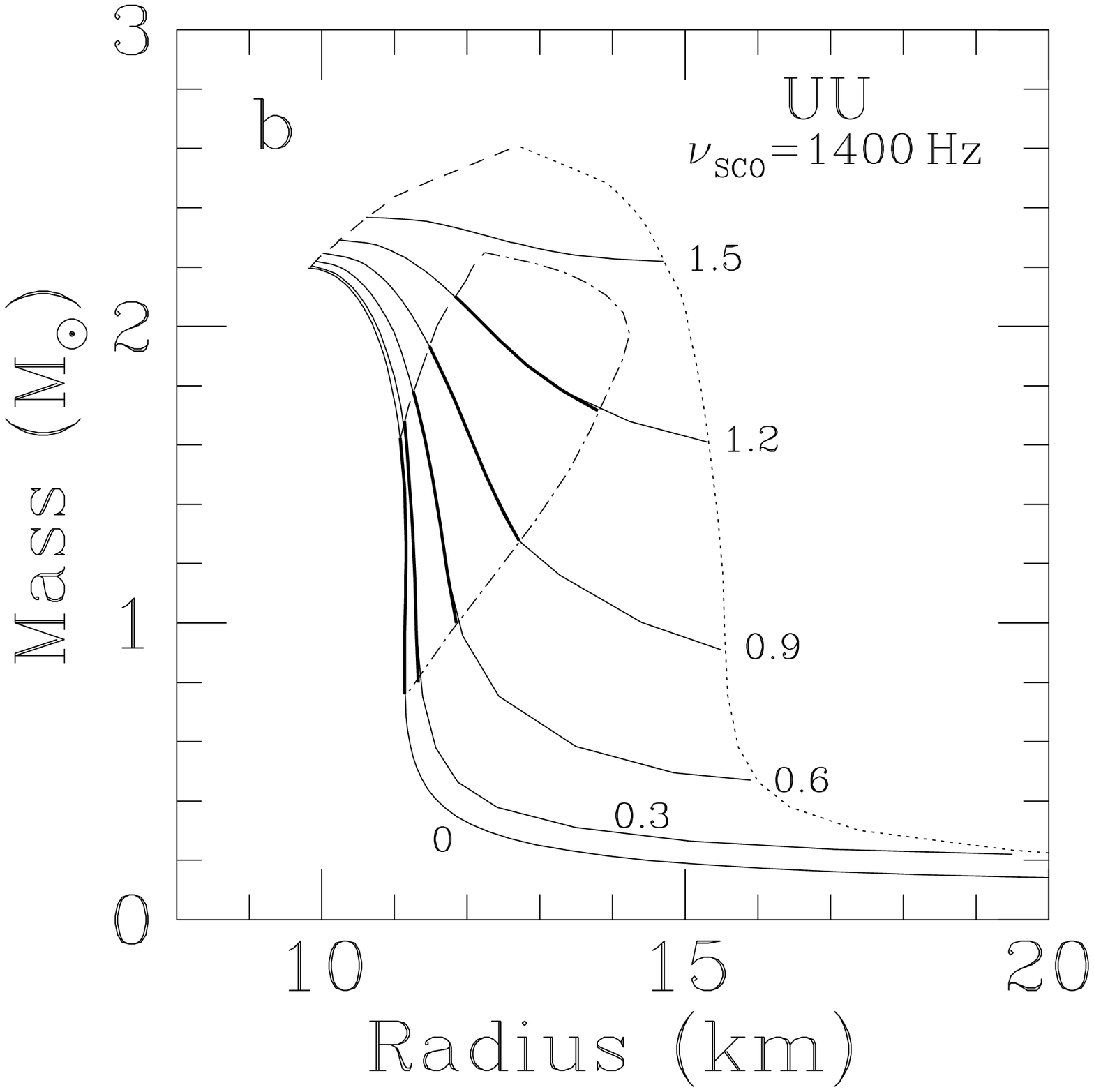,width=3.55truein}
 }
 \caption[fig3]{
 Constraints on neutron star masses and
radii imposed by stellar stability and
observation of a stable circular orbit.
 (a)~Solid lines show the mass-radius
relations for FPS stars with the spin
frequencies indicated (in kHz). The
high-mass end of each mass-radius
relation shown is the radial instability limit
(short-dashed line) whereas the low-mass
end is the mass shedding limit (dotted
line). The dash-dotted line shows the
lowest stellar mass (largest stellar
radius) consistent with the requirement
that the radius of the star be smaller
than the radius of a 1200~Hz orbit. For
FPS stars and this orbital frequency,
the requirement that the radius of the
orbit also exceed the radius of the ISCO
does not constrain the mass or radius of
the star. The bold portion of each mass-radius
curve highlights the region allowed
by both the physical limits of the equation
of state and by the observation of an SCO
with the listed frequency.
 (b)~Same as in (a) but for UU stars and
an orbital frequency of 1400~Hz.
Although the dash-dotted line curves
strongly to the left at high stellar
spin rates, it always shows the smallest
stellar mass (largest stellar radius)
consistent with the requirement that the
radius of the star be smaller than the
radius of a 1400~Hz orbit. The line of
long dashes shows the largest stellar
mass (smallest stellar radius)
consistent with the requirement that the
radius of the orbit exceed the radius of
the ISCO.
 }
 \end{figure*}

\subsection{Mass and Radius Bounds}

As explained in \S~1, observation of a
given SCO frequency around a nonrotating
star allows one to derive upper bounds on
the mass and radius that are independent
of the equation of state, whereas for a
rotating star one must consider a
specific equation of state in order to
derive bounds on the mass and radius.

Given an equation of state and a stellar
spin rate, the mass of the star must be
between the mass-shedding limit and the radial
instability limit,
regardless of the values of any orbital
frequencies. The radius of the star is
bounded by the extreme values of the
radii given by the mass-radius relation
over this mass interval. Observation of
an SCO with a certain frequency {\em
may\/} allow one to restrict {\em
further\/} the allowed mass and radius
intervals, depending on the equation of
state and the frequency of the SCO. The
possible further restrictions are of two
types: a lower bound on the mass,
imposed by the requirement that the
radius of the orbit be greater than the
radius of the star, and an upper bound
on the mass, imposed by the requirement
that the radius of the orbit be equal to
or greater than the radius of the ISCO.
If either of these bounds restrict
further the allowed mass range, the
radius of the star is bounded by the
extreme values of the radii given by the
mass-radius relation over this reduced
mass interval.

Figure~3 shows the constraints on
neutron star masses and radii imposed by
stellar stability and observation of a
stable circular orbit, for stars
constructed using the FPS and UU
equations of state (see \S~2.2). The
mass-radius relations (curves of
gravitational mass versus equatorial
circumferential radius) shown in this
figure were constructed by generating
several sequences of stellar models. Each
sequence consisted of stellar models
with the same baryon number but a range
of spin frequencies. The grid of models
constructed in this way was then used to
generate the mass-radius relations shown
in Figure~3. These relations are
tabulated in Tables~1 and~2, where for each
spin frequency the boundary between normal
and supramassive stars is indicated by the
dashed lines. The relations shown
for rapidly spinning stars are
much flatter than the usual mass-radius
relations for nonrotating stars.

The high-mass end of each constant spin-frequency
sequence shown in Figure~3 is the
gravitational mass above which the star
is unstable to a radial instability.
This mass limit is indicated by the
short-dashed line in Figures~3a and~3b.
The maximum gravitational mass increases
with increasing spin rate, both because
the gravitational mass corresponding to
a given baryonic mass increases with increasing
spin and because the maximum stable baryonic mass
increases with increasing spin. The low-mass 
end of each mass-radius relation shown is the
gravitational mass below which the star
is subject to mass-shedding at the
equator. This mass limit is indicated by
the dotted lines in Figures~3a and~3b.

Consider now the possible further
restrictions on the allowed mass and
radius intervals imposed by observation
of an SCO with a high frequency.
The requirement that the radius of
an orbit be larger than the equatorial
radius of the star places a lower bound
on the mass of the star given the frequency
of an SCO. If, for that
star's spin rate, there are stable stars
with masses smaller than this lower bound,
then the observation of an SCO raises the lower
bound on the mass of a star.
 Whether observation of an SCO with a
given frequency raises the lower bound
on the stellar mass depends on the
equation of state and the star's spin
rate as well as the frequency of the
SCO. For example, for the FPS equation
of state and a spin rate 600~Hz,
observation of a 1200~Hz SCO imposes a
lower bound on the mass of
$0.8\,M_\odot$, whereas the
mass-shedding limit is $0.56\,M_\odot$
(see Fig.~3a). Hence, in this case the
limit imposed by the SCO is stricter.
On the other hand, if $\nu_{\rm spin} >
1200~\Hz$, there is always an SCO with a
frequency of at least 1200~Hz around any
FPS star that is stable against
mass-shedding and hence the SCO
observation does not further restrict
the allowed mass interval. In contrast,
the lower mass limit imposed by
observation of an SCO with a frequency of
1400~Hz is always stricter than the mass
shedding limit for a UU star, regardless
of its spin rate.

Observation of an SCO lowers the upper
bound on the mass of a star if, for that
star's spin rate, there are stable stars
with ISCOs with radii larger than that
permitted by the requirement that the
radius of the orbit be larger than the
radius of the ISCO. Again, whether
observation of an SCO with a given
frequency lowers the upper bound on the
stellar mass depends on the equation of
state and the star's spin rate as well
as the frequency of the SCO. For
example, for the FPS equation of state
observation of a 1200~Hz SCO frequency
would not lower the upper bound on the
stellar mass imposed by the radial
instability limit, regardless of the
star's spin rate. In contrast, for the
UU equation of state observation of a
1400~Hz SCO frequency would lower the
upper bound on the stellar mass imposed
by the radial instability limit,
regardless of the star's spin rate.

Figure~3 shows clearly that observation
of an SCO with a frequency $\gta$1200~Hz
usually reduces greatly the area of the
radius-mass plane allowed for stars
constructed using a given equation of
state. For example, the area allowed for
the UU equation of state if a 1400~Hz
SCO is observed is only a small fraction
of the area allowed by the requirement
of stellar stability (see Fig.~3b).

If the spin frequency of a star is known,
the upper and lower bounds on its mass
and radius imposed by observation of an
SCO of a given frequency can be read off
Figure~3 by looking for the
intersections of the relevant bounding
curves with the mass-radius curve for
that spin frequency. For example, if the
spin frequency is 600~Hz and the SCO
frequency is 1200~Hz, then a neutron star
with the FPS equation of
state must have a mass between
$0.80\,M_\odot$ and $1.82\,M_\odot$ and
a radius between 9.39~km and 12.26~km.

Even if the spin frequency of a neutron
star is unknown, one can still extract
upper and lower bounds on the mass and
radius of the star for a given equation
of state, using the highest observed
frequency of an SCO from the source and
the appropriate figure like Figure~3a
or~3b. For example, if a 1200~Hz SCO is
observed, then a neutron star with the
FPS equation of state must have a mass
between $0.64\,M_\odot$ and
$2.12\,M_\odot$ and a radius between
9.28~km and 15.1~km.

Figures~3a and~3b also show that for a
given equation of state, knowledge of
any two of the mass, radius, or spin
frequency fixes the value of the third
quantity, which may or may not be
consistent with the kilohertz QPO
frequency. Thus, if the spin frequency
of a neutron star in an LMXB is known
and the radius or mass can be determined
by means other than observation of a
kilohertz QPO (e.g., by measuring the
emitting area during a thermonuclear
X-ray burst or by measuring the mass
dynamically), then observation of a
kilohertz QPO will overdetermine the
properties of the star, providing a
check on the consistency of the mass and
radius estimates.

\section{CONCLUSIONS}

Our results show that deviations from
a first-order treatment of the effects of
spin on the structure of neutron stars
and on circular orbits around them are
typically significant for spin frequencies
$\gta400$~Hz. The Kerr spacetime is
generally a poor approximation to the
exterior spacetime of neutron stars
spinning this fast or faster.

Our results demonstrate that the
upper bounds on the stiffness
of neutron star matter implied by
the high frequencies and coherences of
the kilohertz QPOs are tightened
significantly if the star is rotating
rapidly. 
The constraints on the equation of state
become much tighter
if observations conclusively identify a
QPO frequency as the orbital frequency at
the ISCO. For example, if the orbital frequency
at the ISCO is 1100~Hz, the required mass
is $2.0\msun$ for nonrotating stars and
substantially more for rotating stars. Such
a high mass would rule out many of the
softer equations of state and would imply
that the three-nucleon interaction is
strongly repulsive at high
densities (see Pandharipande et al.\ 1998).

As shown in \S~3.3, even if the spin rate
of the star is unknown, the area of the
mass-radius plane allowed by observation
of a high-frequency SCO can be quite
small. If the spin rate is known, the
range of radii allowed for a given equation
of state is usually very small.

Results of the kind presented in this work
will be still more constraining if
measurements of SCO frequencies can be
combined with constraints on other
quantities, such as the stellar
compactness $M/R$ (see Strohmayer 1997;
Miller \& Lamb 1998; Lamb et al.\ 1998;
Strohmayer et al.\ 1998) or the radius of
the star (see Strohmayer et al.\ 1997b;
Strohmayer et al.\ 1998).

Additional observations of kilohertz QPO
sources are extremely important, because
these observations could provide very
tight and robust constraints on the
fundamental properties of neutron stars
and on the equation of state of neutron
star matter.

\acknowledgements

This work was supported in part by NSF
grant AST~96-18524, NASA grant
NAG~5-2925, and NASA RXTE grants at the
University of Illinois, NASA grant
NAG~5-2868 at the University of Chicago,
and through the \gro Fellowship Program,
by NASA grant NAS~5-2687.

\newpage

\begin{table*}[t]
\centering
\begin{tabular}{ccccccccccccccccc}\hline
\multicolumn{17}{c}{\bfseries Table 1:
Mass-Radius Relations for the FPS Equation of State at Different
Spin Rates}{\rule[-2mm]{0mm}{6mm}}\\
\hline
\multicolumn{17}{c}{ }\\

\multicolumn{2}{c}{0 Hz}&&
\multicolumn{2}{c}{300 Hz}&&
\multicolumn{2}{c}{600 Hz}&&
\multicolumn{2}{c}{900 Hz}&&
\multicolumn{2}{c}{1200 Hz}&&
\multicolumn{2}{c}{1300 Hz}\\
\cline{1-2}\cline{4-5}\cline{7-8}\cline{10-11}\cline{13-14}\cline{16-17}
M&R&&M&R&&M&R&&M&R&&M&R&&M&R\\

0.096&50.79&&0.22&19.2 &&0.486&16.64
&&0.94 &15.6 &&1.54 &15.1 &&1.69 &15.0 \\
0.117&26.38&&0.26&15.3 &&0.51 &15.2
&&0.95 &15.2 &&1.56 &14.51&&1.74 &13.4 \\
0.142&19.54&&0.30&13.9 &&0.56 &13.7
&&0.97 &14.8 &&1.60 &13.53&&1.79 &12.6 \\
0.170&16.30&&0.34&13.1 &&0.64 &13.0
&&1.02 &14.2 &&1.68 &12.70&&1.82 &12.15\\
0.203&14.50&&0.41&12.4 &&0.72 &12.6
&&1.14 &13.1 &&1.73 &12.2 &&1.835&11.94\\
0.242&13.39&&0.48&12.0 &&0.80 &12.3
&&1.25 &12.6 &&1.78 &11.74&&1.842&11.87\\
0.286&12.68&&0.56&11.8 &&0.87 &12.1
&&1.34 &12.2 &&1.81 &11.45&&-- -- --&-- -- --\\
0.336&12.20&&0.64&11.6 &&1.01 &11.9
&&1.41 &12.0 &&1.825&11.31&&1.848&11.80\\
0.392&11.87&&0.72&11.5 &&1.13 &11.7
&&1.48 &11.85&&1.832&11.23&&1.851&11.67\\
0.456&11.65&&0.80&11.4 &&1.24 &11.5
&&1.54 &11.67&&-- -- --&-- -- --&&1.857&11.64\\
0.527&11.49&&0.87&11.4 &&1.33 &11.4
&&1.58 &11.52&&1.837&11.16&&1.873&11.36\\
0.607&11.38&&1.01&11.3 &&1.41 &11.2
&&1.66 &11.2&& 1.841&11.07&&1.889&11.09\\
0.693&11.30&&1.13&11.2 &&1.47 &11.14
&&1.71 &10.99&&1.847&11.05&&1.902&10.73\\
0.785&11.24&&1.24&11.1 &&1.53 &11.04
&&1.76 &10.7&& 1.864&10.81&&1.912&9.995\\
0.882&11.20&&1.33&11.0 &&1.57 &10.92
&&1.79 &10.49&&1.879&10.52&&$\ldots$&$\ldots$\\
0.980&11.15&&1.40&10.9 &&1.65 &10.7
&&1.81 &10.36&&1.892&9.851&&$\ldots$&$\ldots$\\
1.078&11.11&&1.47&10.8 &&1.70 &10.51
&&1.815&10.30&&$\ldots$&$\ldots$&&$\ldots$&$\ldots$\\
1.173&11.05&&1.52&10.74&&1.75 &10.25
&&-- -- --&-- -- --&&$\ldots$&$\ldots$&&$\ldots$&$\ldots$\\
1.264&10.98&&1.57&10.66&&1.78 &10.05
&&1.82&10.21&&$\ldots$&$\ldots$&&$\ldots$&$\ldots$\\
1.349&10.90&&1.64&10.5 &&1.80 &9.90
&& 1.825&10.15&&$\ldots$&$\ldots$&&$\ldots$&$\ldots$\\
1.427&10.81&&1.70&10.29&&1.805&9.82
&& 1.829&10.09&&$\ldots$&$\ldots$&&$\ldots$&$\ldots$\\
1.497&10.71&&1.75&10.04&&-- -- --&-- -- --
&&1.846&9.547&&$\ldots$&$\ldots$&&$\ldots$&$\ldots$\\
1.559&10.60&&1.78&9.83&& 1.811&9.71
&&$\ldots$&$\ldots$&&$\ldots$&$\ldots$&&$\ldots$&$\ldots$\\
1.614&10.47&&1.79&9.65&& 1.815&9.61
&&$\ldots$&$\ldots$&&$\ldots$&$\ldots$&&$\ldots$&$\ldots$\\
1.661&10.34&&1.80&9.54&& 1.821&9.39
&&$\ldots$&$\ldots$&&$\ldots$&$\ldots$&&$\ldots$&$\ldots$\\
1.701&10.20&&-- -- --&-- -- --&&$\ldots$&$\ldots$
&&$\ldots$&$\ldots$&&$\ldots$&$\ldots$&&$\ldots$&$\ldots$\\
1.734&10.05&&1.814&9.39&&$\ldots$&$\ldots$
&&$\ldots$&$\ldots$&&$\ldots$&$\ldots$&&$\ldots$&$\ldots$\\
1.759&9.898&&$\ldots$&$\ldots$&&$\ldots$&$\ldots$
&&$\ldots$&$\ldots$&&$\ldots$&$\ldots$&&$\ldots$&$\ldots$\\
1.778&9.746&&$\ldots$&$\ldots$&&$\ldots$&$\ldots$
&&$\ldots$&$\ldots$&&$\ldots$&$\ldots$&&$\ldots$&$\ldots$\\
1.790&9.594&&$\ldots$&$\ldots$&&$\ldots$&$\ldots$
&&$\ldots$&$\ldots$&&$\ldots$&$\ldots$&&$\ldots$&$\ldots$\\
1.797&9.444&&$\ldots$&$\ldots$&&$\ldots$&$\ldots$
&&$\ldots$&$\ldots$&&$\ldots$&$\ldots$&&$\ldots$&$\ldots$\\
1.799&9.295&&$\ldots$&$\ldots$&&$\ldots$&$\ldots$
&&$\ldots$&$\ldots$&&$\ldots$&$\ldots$&&$\ldots$&$\ldots$\\
\hline

\end{tabular}
\end{table*}
\newpage

\begin{table*}[t]
\centering
\begin{tabular}{ccccccccccccccccc}\hline
\multicolumn{17}{c}{\bfseries Table 2:
Mass-Radius Relations for the UU Equation of State at Different
Spin Rates}{\rule[-2mm]{0mm}{6mm}}\\
\hline
\multicolumn{17}{c}{ }\\

\multicolumn{2}{c}{0 Hz}&&
\multicolumn{2}{c}{300 Hz}&&
\multicolumn{2}{c}{600 Hz}&&
\multicolumn{2}{c}{900 Hz}&&
\multicolumn{2}{c}{1200 Hz}&&
\multicolumn{2}{c}{1500 Hz}\\
\cline{1-2}\cline{4-5}\cline{7-8}\cline{10-11}\cline{13-14}\cline{16-17}
M&R&&M&R&&M&R&&M&R&&M&R&&M&R\\

0.135&21.17&&0.220&19.50&&0.470&15.90
&&0.910&15.50&&1.610&15.30&&2.220&14.70\\
0.151&18.24&&0.231&16.83&&0.480&15.08
&&0.965&14.54&&1.664&14.16&&2.225&14.01\\
0.172&16.06&&0.265&15.07&&0.553&13.32
&&1.167&13.02&&1.777&13.32&&2.242&13.40\\
0.199&14.49&&0.299&13.31&&0.751&12.27
&&1.347&12.53&&1.873&12.77&&2.256&13.10\\
0.232&13.37&&0.377&12.27&&0.959&11.88
&&1.507&12.24&&1.955&12.43&&2.265&12.89\\
0.273&12.58&&0.462&11.82&&1.159&11.72
&&1.641&12.01&&2.021&12.18&&2.274&12.81\\
0.320&12.05&&0.552&11.56&&1.339&11.63
&&1.754&11.83&&2.073&11.96&&2.278&12.69\\
0.376&11.69&&0.749&11.34&&1.498&11.53
&&1.851&11.65&&2.115&11.77&&2.282&12.64\\
0.440&11.45&&0.956&11.30&&1.631&11.41
&&1.933&11.48&&2.148&11.61&&2.284&12.60\\
0.511&11.30&&1.156&11.28&&1.743&11.30
&&1.999&11.34&&2.174&11.49&&-- -- --&-- -- --\\
0.593&11.21&&1.335&11.25&&1.840&11.19
&&2.051&11.20&&2.194&11.38&&2.286&12.56\\
0.686&11.16&&1.493&11.20&&1.921&11.09
&&2.093&11.10&&2.209&11.29&&2.299&12.36\\
0.789&11.14&&1.626&11.16&&1.986&11.00
&&2.126&10.98&&2.220&11.22&&2.312&12.17\\
0.902&11.14&&1.738&11.09&&2.038&10.87
&&2.152&10.89&&2.228&11.18&&2.324&12.02\\
1.022&11.15&&1.834&11.00&&2.080&10.77
&&2.172&10.79&&2.233&11.13&&2.334&11.80\\
1.147&11.15&&1.915&10.90&&2.113&10.67
&&2.187&10.72&&2.240&11.08&&2.345&11.66\\
1.271&11.16&&1.980&10.80&&2.139&10.58
&&2.198&10.66&&-- -- --&-- -- --&&2.354&11.45\\
1.394&11.15&&2.032&10.71&&2.159&10.49
&&2.206&10.61&&2.256&10.95&&2.362&11.24\\
1.513&11.12&&2.073&10.60&&2.174&10.42
&&2.212&10.57&&2.269&10.82&&2.366&10.80\\
1.624&11.08&&2.106&10.52&&2.185&10.36
&&2.219&10.52&&2.275&10.75&&2.367&10.61\\
1.726&11.02&&2.132&10.42&&2.193&10.30
&&-- -- --&-- -- --&&2.281&10.67&&$\ldots$&$\ldots$\\
1.821&10.94&&2.152&10.34&&2.199&10.26
&&2.235&10.39&&2.288&10.38&&$\ldots$&$\ldots$\\
1.907&10.85&&2.167&10.26&&2.206&10.19
&&2.247&10.01&&2.291&10.23&&$\ldots$&$\ldots$\\
1.980&10.74&&2.179&10.19&&-- -- --&-- -- --
&&$\ldots$&$\ldots$&&$\ldots$&$\ldots$&&$\ldots$&$\ldots$\\
2.041&10.63&&2.186&10.12&&2.207&10.16
&&$\ldots$&$\ldots$&&$\ldots$&$\ldots$&&$\ldots$&$\ldots$\\
2.091&10.51&&2.190&10.06&&2.214&10.04
&&$\ldots$&$\ldots$&&$\ldots$&$\ldots$&&$\ldots$&$\ldots$\\
2.129&10.38&&2.199& 9.96&&2.219&9.92
&&$\ldots$&$\ldots$&&$\ldots$&$\ldots$&&$\ldots$&$\ldots$\\
2.157&10.25&&-- -- --&-- -- --&&$\ldots$&$\ldots$
&&$\ldots$&$\ldots$&&$\ldots$&$\ldots$&&$\ldots$&$\ldots$\\
2.177&10.13&&2.205&9.86&&$\ldots$&$\ldots$
&&$\ldots$&$\ldots$&&$\ldots$&$\ldots$&&$\ldots$&$\ldots$\\
2.189&10.01&&$\ldots$&$\ldots$&&$\ldots$&$\ldots$
&&$\ldots$&$\ldots$&&$\ldots$&$\ldots$&&$\ldots$&$\ldots$\\
2.195&9.898&&$\ldots$&$\ldots$&&$\ldots$&$\ldots$
&&$\ldots$&$\ldots$&&$\ldots$&$\ldots$&&$\ldots$&$\ldots$\\
2.196&9.814&&$\ldots$&$\ldots$&&$\ldots$&$\ldots$
&&$\ldots$&$\ldots$&&$\ldots$&$\ldots$&&$\ldots$&$\ldots$\\
\hline

\end{tabular}
\end{table*}

\begin{references}

\reference{APR98}
 Akmal, A., Pandharipande, V.~R., \&
Ravenhall, D.~G. 1998, Phys. Rev. C. 
submitted (nucl-th/9804027)

\reference{AB77}
 Arnett, W.\,D., \& Bowers, R.\,L. 1977, ApJ
Suppl., 33, 415

\reference{CST92}
 Cook, G. B., Shapiro, S. L., \&
Teukolsky, S. A. 1992, ApJ, 398, 203

\reference{CST94a}
 ---------. 1994a, ApJ, 422, 227

\reference{CST94b}
 ---------. 1994b, ApJ, 424, 823

\reference{FP82}
 Friedman, B., \& Pandharipande, V.\,R.
1981, Nucl. Phys. A, 361, 501

\reference{G92}
 Glendenning, N.\,K. 1992, Phys. Rev. D, 
46, 1274

\reference{HT68}
 Hartle, J.\,B., \& Thorne, K. S. 1968,
ApJ, 153, 807

\reference{HPS93}
 Heiselberg, H., Pethick, C.\,J., \&
Staubo, E.\,F. 1993, Phys. Rev. Lett.,
70, 379

\reference{KMW1990}
 Klu\'zniak, W., Michelson, P., \&
Wagoner, R.\,V. 1990, ApJ, 358, 538

\reference{KEH89a}
 Komatsu, H., Eriguchi, Y., \& Hachisu,
I. 1989a, MNRAS, 237, 355

\reference{KEH89b}
 ---------. 1989b, MNRAS, 239, 153

\reference{LP81}
 Lagaris, I.\,E., \& Pandharipande,
V.\,R. 1981, Nucl. Phys., A359, 349

\reference{LMP98}
 Lamb, F.\,K., Miller, M.\,C., \&
Psaltis, D. 1998, in Nuclear
Astrophysics, Proc. International
Workshop XXVI on Gross Properties of
Nuclei and Nuclear Excitations, ed. M.
Buballa, N. N\"orenberg, J. Wambach, \&
A. Wirzba (Darmstadt: GSI), 114
(astro-ph/9802348)

\reference{Lightman73}
 Lightman, A.\,P., Press, W.\,H., Price,
R.\,H., \& Teukolsky, S.\,A. 1973,
Problem Book in Relativity and
Gravitation (Princeton: Princeton
University Press)

\reference{LRP93}
 Lorenz, C.\,P., Ravenhall, D.\,G., \&
Pethick, C.\,J. 1993, Phys. Rev. Lett.
70, 379

\reference{ML93}
 Miller, M.\,C., \& Lamb, F.\,K. 1993,
ApJ, 413, L43

\reference{ML96}
 ---------. 1996, ApJ, 470, 1033

\reference{ML98} ---------.
 1998, ApJ Lett., in press
(astro-ph/9711325)

\reference{MLP98a}
 Miller, M.\,C., Lamb, F.\,K., \& Psaltis,
D. 1998a, ApJ, in press

\reference{MLP98b}
 ---------. 1998b, in preparation

\reference{PAR98}
 Pandharipande, V.\,R., Akmal, A., \&
Ravenhall, D.\,G. 1998, in Nuclear
Astrophysics, Proc. International Workshop
XXVI on Gross Properties of Nuclei and
Nuclear Excitations, ed. M. Buballa, N.
N\"orenberg, J. Wambach, \& A. Wirzba
(Darmstadt: GSI), 11

\reference{PS75a}
 Pandharipande, V.\,R., \& Smith, R.\,A.
1975a, Nucl. Phys., A237, 507

\reference{PS75b}
 ---------. 1975b, Phys. Letters, 59B, 15

\reference{PR95}
 Pethick, C.\,J., \& Ravenhall, D.\,G.
1995, Ann. Rev. Nucl. Par. Sci., 45, 429

\reference{SPW86}
 Schiavilla, R., Pandharipande, V.\,R.,
\& Wiringa, R.\,B. 1986, Nucl. Phys.,
A449, 219

\reference{Smith97}
 Smith, D.\,A., Morgan, E.,\,H., \&
Bradt, H. 1997, ApJ, 479, L137

\reference{Stroh97}
 Strohmayer, T.\,E. 1997, talk presented
at the 1997 meeting of the High Energy
Astrophysics Division of the American
Astronomical Society in Estes Park,
Colorado

\reference{Stroh97a}
 Strohmayer, T.\,E., Jahoda, K., Giles,
A.\,B., \& Lee, U. 1997a, ApJ, 486, 355

\reference{SZS97}
 Strohmayer, T.\,E., Zhang, W., \& Swank,
J.\,H. 1997b, ApJ, 487, L77

\reference{Stroh96}
 Strohmayer, T., Zhang, W., Swank, J.\,H.,
Smale, A., Titarchuk, L., \& Day, C.
1996, ApJ, 469, L9

\reference{Stroh98}
 Strohmayer, T.\,E., Zhang, W., Swank,
J.\,H., White, N.\,E., \& Lapidus, I.
1998, ApJ Lett., in press
(astro-ph/9803119)

\reference{Swank97}
 Swank, J. 1997, talk presented at the
Symposium on The Active X-ray Sky:
Results from BeppoSAX and Rossi-XTE,
Accademia dei Lincei, Rome, October 1997

\reference{vdK95} van der Klis, M. 1995, in X-Ray
Binaries, ed. W.\,H.\,G. Lewin, J.\ van Paradijs,
\& E.\,P.\,J. van den Heuvel (Cambridge U.\,P.), 252

\reference{vdK98}
 ---------. 1998, in The Many Faces
of Neutron Stars, Proc.\ NATO ASI,
Lipari, Italy, in press
(astro-ph/9710016)

\reference{Wij98}
Wijnands, R., M{\'e}ndez, M., van der Klis,
M., Psaltis, D., Kuulkers, E., \& Lamb, F.\
K.\ 1998, \apj, submitted

\reference{WK97}
 Wijnands, R.\,A.\,D., \& van der Klis,
M. 1997, ApJ, 482, L65

\reference{WK98a}
 ---------. 1998a, IAUC 6876
 
\reference{WK98b}
 ---------. 1998b, submitted to Nature
(astro-ph/9804216)

\reference{W97}
 Wijnands, R.\,A.\,D., van der Klis, M.,
van Paradijs, J., Lewin, W.\,H.\,G.,
Lamb, F.\,K., Vaughan, B., \& Kuulkers,
E. 1997, ApJ, 479, L141

\reference{WFF88}
 Wiringa, R.\,B., Fiks, V., \& Fabrocini,
A. 1988, Phys. Rev. C, 38, 1010

\reference{Zhang97}
 Zhang, W., Lapidus, I., Swank, J.\ H.,
White, N.\ E., \& Titarchuk, L.\ 1997,
IAU Circ. 6541


\end{references}
\end{document}